\documentclass[pdflatex,sn-mathphys-ay]{sn-jnl}

\usepackage{graphicx}%
\usepackage{multirow}%
\usepackage{amsmath,amssymb,amsfonts}%
\usepackage{amsthm}%
\usepackage{mathrsfs}%
\usepackage[title]{appendix}%
\usepackage{xcolor}%
\usepackage{textcomp}%
\usepackage{manyfoot}%
\usepackage{booktabs}%
\usepackage{algorithm}%
\usepackage{algorithmicx}%
\usepackage{algpseudocode}%
\usepackage{listings}%

\usepackage{dirtytalk}
\usepackage{multirow}
\DeclareMathOperator*{\rint}{\ThisStyle{\rotatebox{15}{$\SavedStyle\!\int\!$}}}
\usepackage{scalerel}
\usepackage{graphicx}
\usepackage{yfonts}
\usepackage{booktabs}

\theoremstyle{thmstyleone}%
\theoremstyle{thmstyletwo}%

\theoremstyle{thmstylethree}%

\begin{document}

\title[Method for restoring the orientation of ocean bottom seismometers using distant earthquakes records]{Method for restoring the orientation of ocean bottom seismometers using distant earthquakes records}


\author[1]{\fnm{Oleg V.} \sur{Ponomarev}}\email{ponomarev.ov20@physics.msu.ru}

\author[1,2]{\fnm{Sergey V.} \sur{Kolesov}}\email{kolesov@phys.msu.ru}

\author[1,2]{\fnm{Mikhail A.} \sur{Nosov}}\email{m.a.nosov@mail.ru}

\affil[1]{\orgdiv{Department of Physics}, \orgname{Lomonosov Moscow State University}, \orgaddress{\city{Moscow}, \postcode{119991}, \country{Russia}}}

\affil[2]{\orgname{Institute of Marine Geology and Geophysics, Far Eastern Branch, Russian Academy of Science}, \orgaddress{\city{Yuzhno-Sakhalinsk}, \postcode{693022}, \country{Russia}}}


\abstract{A method is presented for determining the vertical direction relative to the axes of seismometers installed in seafloor observatories. The method is based on the linear relationship between the vertical component of seafloor acceleration and pressure variations at the ocean bottom, which follows directly from Newton’s second law and holds within the frequency range of \say{forced oscillations}. The method’s performance was validated using data from ocean bottom seismometers (accelerometers) and pressure gauges of six S-net stations during three seismic events: Gulf of Alaska $M_w 7.9$ (23.01.2018), Chignik $M_w 8.2$ (29.07.2021) and Noto $M_w 7.5$ (01.01.2024). The results were compared with an independent alternative method, showing discrepancies within $1^\circ$ for most stations. A key advantage of the proposed method is its applicability not only to accelerometers but as well to velocimeters.}




\maketitle

\section{Introduction}\label{sec1}

The first experiments involving seismic measurements on the ocean floor date back to the first half of the 20th century~\cite{1}, while active development of this field began much later --- in the 1960s~\cite{bradner}. Around the same time, a hydrophysical method for tsunami forecasting was proposed and implemented, based on high precision measurements of pressure variations on the seafloor~\cite{sol1, sol2}.\par

In the early 21st century, the deployment of ocean bottom seismometers (OBS) and pressure gauges (PG) became widespread~\cite{Suetsugu, 2.3, 14}. One of the most well-known continuously operating systems is DART (deep-ocean assessment and reporting of tsunamis, https://nctr.pmel.noaa.gov/Dart/), which consists of approximately 60 autonomous deep-ocean tsunami-meters located at various locations in the World ocean, transmitting sea level information via the hydroacoustic channel and satellite communication~\cite{mungov}. Compared to autonomous stations, cabled systems have clear advantages in terms of data volume, transmission speed, and operational lifespan. The largest cabled systems currently in operation are DONET (Dense Oceanfloor Network System for Earthquakes and Tsunamis — 51 stations)~\cite{2} and S-net (150 stations)~\cite{3}, both installed in the Pacific Ocean near the Japanese islands. In addition, there are many other systems, such as NEPTUNE (The Canadian North-East Pacific Underwater Networked Experiments)~\cite{barnes}, EMSO (European Multidisciplinary Seafloor and Water Column Observatory)~\cite{favali}, MACHO (MArine Cable Hosted Observatory)~\cite{hsiao}, among others.\par

Seafloor observatories enable the resolution of numerous important scientific and practical tasks, the foremost of which is obtaining the information about underwater earthquakes and/or tsunamis in close proximity to the source, i.e., in the shortest possible time.\par

The characteristics of signals recorded by OBS and PG during seismic seafloor movements were discussed as early as the late 20th century in works such as~\cite{filloux, webb}. A comprehensive physical model, based on advanced theoretical understanding and analysis of synchronous measurements of pressure variations and vertical accelerations of the seafloor, is presented in~\cite{bol, 14, pre10}. A compressible water layer of depth $H$ is characterized by two characteristic frequencies: $f_g=0.366 \cdot \sqrt{g/H}$ and $f_{ac}=c/(4H)$, where $c$ is the speed of sound in water and $g$ is gravitational acceleration. Seismic motions of the seafloor with frequency $f<f_g$ generate gravity waves, whereas motions with $f>f_{ac}$ generate hydroacoustic waves. If the seafloor moves within the frequency range $f_g<f<f_{ac}$, neither gravity waves nor hydroacoustic waves are generated, and the water layer undergoes forced oscillations that replicate the motion of the seafloor. In the case of a flat horizontal seafloor, a relationship holds between bottom pressure variations $p$ and the vertical acceleration component $a_v$, which follows directly from Newton's second law:

\begin{equation}
p=\rho H \cdot a_v, \label{2nl}
\end{equation}
where $\rho$ is the vertically averaged water density. The validity of Equation~\ref{2nl} has been confirmed in a number of studies~\cite{an, matsu, 15, 11, deng}.\par

The strict fulfillment of the relationship~\ref{2nl} enabled the development of the Method for Examining the Performance of Seafloor Observatory Sensors, and in particular, allowed the identification of incorrect calibration of station E18 (DONET)~\cite{pre10, 10}.\par

The practical applications of the relationship~\ref{2nl} are not limited to verifying sensor calibration. It is known that strong, nearby earthquakes can alter the orientation of seismometer axes. Considering the S-net system, the seismometer axes are not aligned with any fixed directions at all~\cite{9}. There exists a method for recovering accelerometer axis orientation based on detecting the constant components of acceleration signals due to gravity~\cite{12}. However, this method is not universal; for example, it cannot be applied to the recovery of velocity sensor axes.\par

The aim of this study is to develop a method for determining the vertical direction relative to the axes of an ocean bottom seismometer based on the relationship~\ref{2nl}. The functionality of the method is demonstrated using data from six S-net stations for the following three earthquakes: Gulf of Alaska $M_w 7.9$ (23.01.2018), Chignik $M_w 8.2$ (29.07.2021), and Noto $M_w 7.5$ (01.01.2024).

\section{Method Description}\label{sec2}

Seismometers typically have three orthogonal axes: $x$, $y$, and $z$, oriented toward the east, north, and vertically upward, respectively. However, the actual orientation of these axes may deviate due to errors during the installation of observatories or shifts occurring during operation, caused by strong seismic motions or other factors. The analyzed dataset includes time series of pressure and the three mutually perpendicular acceleration components. \par

According to equation~\ref{2nl}, the application of the method requires the presence of some nonzero dynamic acceleration at the seafloor; therefore, data from periods of earthquake activity are used. It is known that the ability of the earthquake to excite low-frequency seismic waves grows with an increase in magnitude $M_w$~\cite{mag}. Therefore, it is important to select earthquakes that are strong enough to produce ocean bottom motions within the range of forced oscillations (for stations at depths $H \in [1; \; 3] \; km$, $M_w \geq 7.5$~\cite{11}).\par

Until this point, the seafloor was assumed to be flat and horizontal. The case of sloped bathymetry was discussed in detail in~\cite{pre10}: under real seafloor conditions, horizontal acceleration components also begin to influence pressure variation. However, it was shown that, due to the generally small depth gradients in the real ocean, the normal to the seafloor surface can be assumed to be close to the vertical direction, and the influence of steep underwater slopes can be neglected if they are located outside the range with a radius of $3H$. \par

We will define a Cartesian coordinate system $0xyz$ associated with the seismometer. Let the accelerations along the three orthogonal axes be known: $a_x(t)$, $a_y(t)$, $a_z(t)$. Then, the projection of acceleration onto an arbitrary spatial direction can be defined using two angles, $\eta$ and $\kappa$:

\begin{equation}
    a_{\eta, \kappa}(t) = a_z(t) \cdot cos(\eta) + a_y(t) \cdot sin(\eta) \cdot sin(\kappa) + a_x(t) \cdot sin(\eta) \cdot cos(\kappa)
\end{equation}
where $\eta \in [0, 180]$ is the angle between the selected direction and the $0z$ axis, and $\kappa \in [0, 360)$ is the angle between the projection of the selected direction onto the $0xy$ plane and the $0x$ axis. \par

Evidently, a certain linear combination of measured acceleration components $a_x, a_y, a_z$, given by a pair of angle values $\eta, \kappa$ will represent acceleration along the vertical direction $a_v$. The algorithm for determining these angles is based on cross-spectral analysis and consists of three stages. \par

In the \textit{first stage}, the pressure variation is computed as $p = P - \overline{P}$, where $P$ is the measured pressure at the seafloor, and $\overline{P}$ is the approximation of the constant pressure component and tidal trend using an 8th-degree polynomial. \par

In the \textit{second stage}, a set of acceleration projections is generated along vectors corresponding to different combinations of angles $\eta \in [0, 180]$ and $\kappa \in [0, 360)$, iterated with a certain step: e.g., $1^\circ$. For each pair $\eta, \kappa$, the MSC (magnitude-squared coherence) is computed between the pressure variation $p$ and the acceleration projection $a_{\eta, \kappa}(t)$. The pressure and acceleration data must be resampled to the same frequency. In the ideal case, the MSC value calculated for angle values $\eta, \kappa$ corresponding to the vertical direction within the frequency range $f_{g} < f < f_{ac}$ should be close to unity. Examples of MSC results computed using data from 10 DONET stations during the 2011 Tohoku earthquake are presented in~\cite{pre10}. \par

In the \textit{third stage}, the coherence integral $I$ is calculated for each direction:

\begin{equation}
    I = \rint_{f_g}^{f_{up}} MSC(f) \,df,
\end{equation}
where the upper frequency limit $f_{up}$ is defined by:
\begin{equation}    
    f_{up} = min(f_{ac}, \; 0.1Hz) \label{fup}
\end{equation}

As demonstrated in~\cite{11}, microseismic noise may occur at frequencies above 0.1 Hz, which can disrupt the validity of equation~\ref{2nl}. For this reason, when computing the coherence integral $I$, the upper frequency limit is defined by expression~\ref{fup} rather than simply using $f_{ac}$. The integral value at each point is normalized as follows:

\begin{equation}
    I_{norm} = \frac{I - I_{min}}{I_{max}- I_{min}},
\end{equation}

where $I_{min}$ and $I_{max}$ are the minimum and maximum values of the integral computed across all directions. Consequently, the values of the integral plotted on the diagram will always lie in the range from $0$ to $1$, where the maximum value corresponds to an ideal linear relationship between the pressure variation and the acceleration projection onto the selected vector, while decreasing values indicate a weakening correlation. Due to the relation~\ref{2nl}, the value of $I$ will be maximized for the angles corresponding to the vertical direction. To aid visualization, a diagram of $I_{norm}(\eta, \kappa)$ values is constructed. Since the method is based on magnitude-squared coherence, the vertical direction can only be determined up to a sign. According to the theory outlined above, the diagram will exhibit two maxima at diametrically opposite points corresponding to the directions of $\vec{a}_v$ and $-\vec{a}_v$. The coordinates of these two maxima in the $(\eta, \kappa)$ plane are related by the expressions: $\eta_2 = \pi - \eta_1, \kappa_2 = \pi + \kappa_1$. In the ideal case, where the $z$-axis of the accelerometer aligns with the vertical, the diagram should show two strip-like maxima aligned along $\eta = 0^\circ$ and $\eta = 180^\circ$. \par

The core assumption of this article is that the linear combination of the measured accelerations $a_x, a_y, a_z$, which yields the maximum coherence integral value $I_{max}$, corresponds to the acceleration in the vertical direction: $a_{I_{max}}(\eta, \kappa, t) = a_v(t)$. Applying this method results in a pair of angle values $\eta, \kappa$ that define the vertical direction relative to the seismometer's axes.

\section{Data and their processing}\label{sec3}

The method described above is applied to several stations from the S-net network~\cite{snet, aoi}. The network comprises 150 seafloor observatories, conventionally divided into six branches. Branches S1 through S5 run along the eastern coast of Japan, while the sixth branch closes the loop by extending into the open ocean. The station locations are shown in Fig.~\ref{fig:map}b. The first five branches (S1–S5) were commissioned on August 15, 2016, and the sixth branch (S6) became operational in August 2017.\par

As mentioned in the discussion of the method's applicability, the seafloor in the vicinity of a sensor must be sufficiently smooth and free from abrupt depth variations. For each of the 150 stations, depth variation within a radius of $3H$ was evaluated. The variation at a point was defined as the magnitude of the depth gradient:

\begin{equation}
\delta = |grad(H)|
\end{equation}

The resulting value $\delta$ is a scalar, dimensionless quantity that physically corresponds to the tangent of the local seafloor slope at a given point.\par

Seafloor slope maps were generated for all S-net stations. Gradients were computed using the GEBCO 2023 bathymetric dataset~\cite{gebco}, which offers a spatial resolution of $15$ arc-seconds. Only stations located at depths between $2$ and $6 km$ were considered, ensuring a sufficiently wide frequency range for analysis. The broadest frequency band over which the coherence integral can be evaluated is achieved at $f_{ac} = 0.1 Hz$, corresponding to a depth of $H = 3750 m$. As a result, six stations were selected that both fall within this depth range and exhibit the lowest average depth gradient within a radius of $3H$: N.S2N17, N.S6N08, N.S2N18, N.S4N25, N.S4N11, and N.S4N06. The average gradient near each of these stations does not exceed $3.34 \cdot 10^{-2}$. These selected stations are marked with red stars in Fig.~\ref{fig:map}b. Bathymetry and gradient maps in the vicinity of each of the six selected stations are provided in supplements. 

\setcounter{figure}{0}

\begin{figure}[]
    \centering
    \includegraphics[width=14cm]{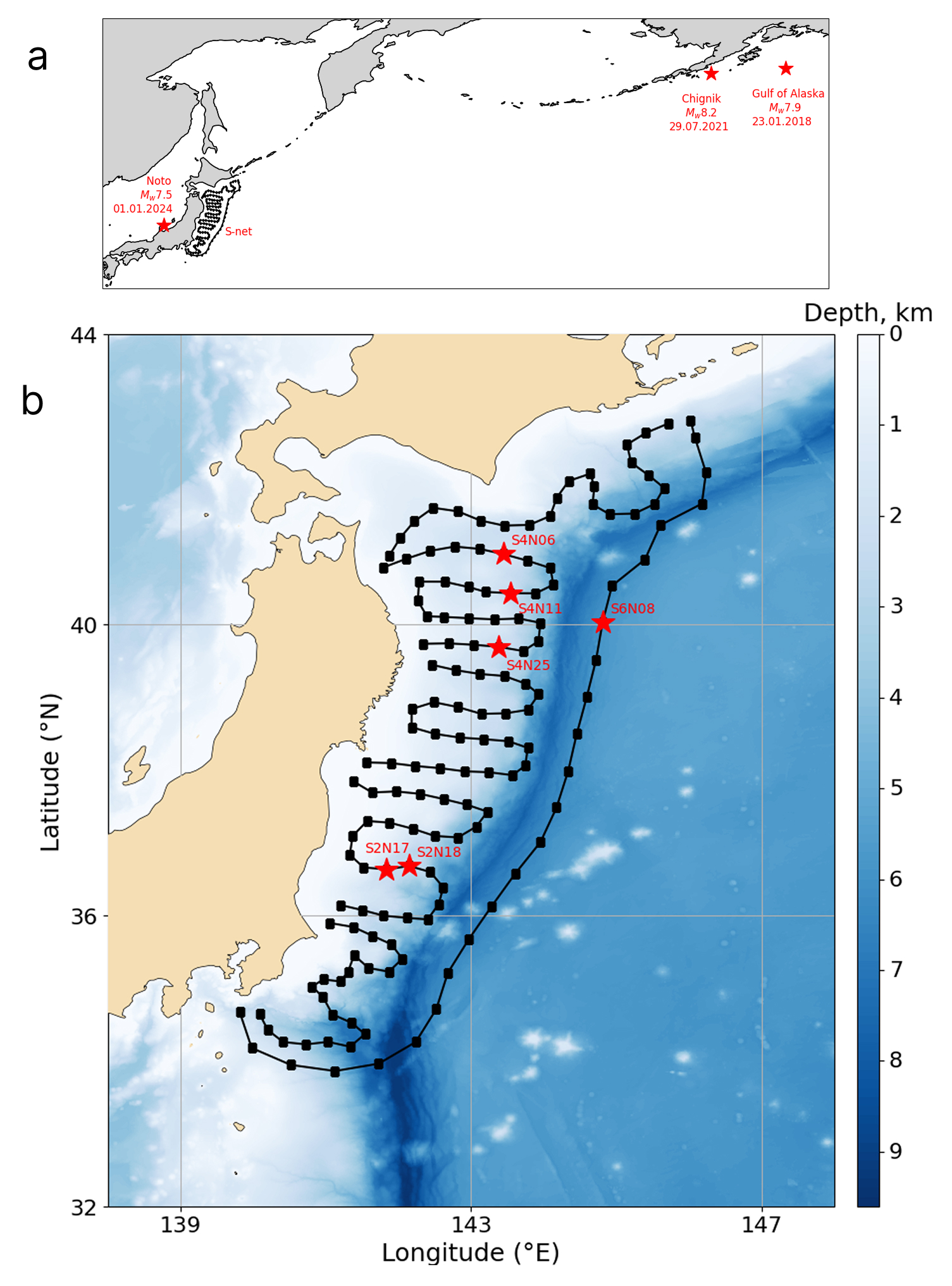}
    \caption{Location of the selected events and the S-net on a map of the North Pacific Ocean. \textbf{(a)}. Detailed map of the S-net stations. \textbf{(b)} Stations selected for analysis are marked with red stars.}
    \label{fig:map}
\end{figure}

As noted above, the S-net network became fully operational in August 2017, and thus seismic events for analysis can be drawn from the period spanning late 2017 to early 2025. In this study, three earthquakes were selected for analysis: Gulf of Alaska ($M_w 7.9$, 23.01.2018), Chignik ($M_w 8.2$, 29.07.2021), and Noto ($M_w 7.5$, 01.01.2024). Their location is shown in Fig.~\ref{fig:map}a. This selection enables a temporal resolution of approximately two years to track potential changes in station orientation. The Noto earthquake is of particular interest due to its proximity to the station locations, in contrast to the other two events, whose epicenters were located thousands of kilometers away: the epicentral distances were $45^\circ$-$49^\circ$ for the Gulf of Alaska, $40^\circ$-$44^\circ$ for Chignik, and $4^\circ$-$6^\circ$ for Noto. This disparity in source distances allows an assessment of how diagram features vary with signal intensity. Notably, the peak pressure amplitude recorded by station N.S2N17 during the Noto event was at least $20$ times higher than that recorded during the Chignik and Gulf of Alaska events.\par

\begin{figure}[H]
    \centering
    \includegraphics[width=14cm]{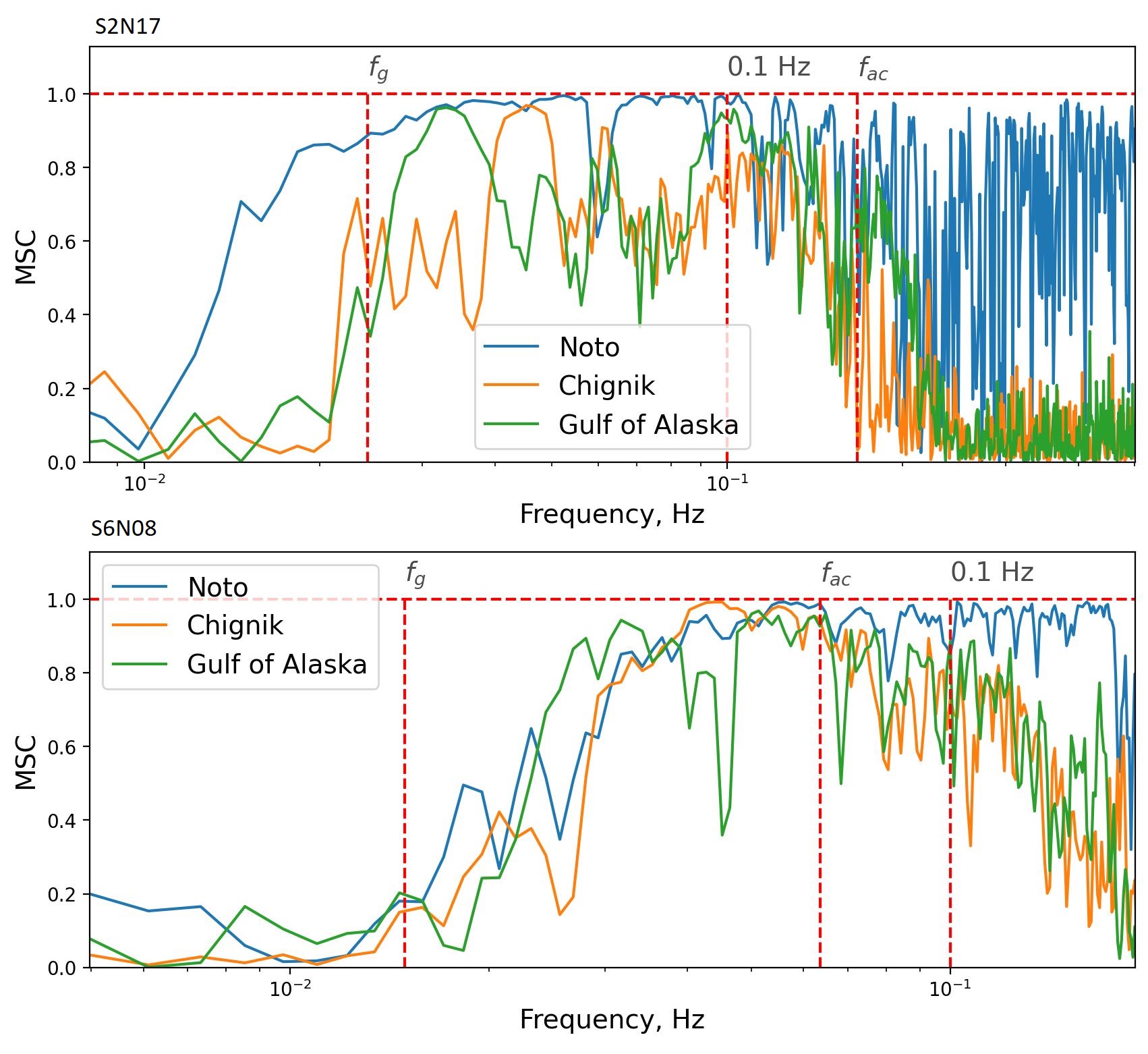}
    \caption{Cross-spectra (magnitude-squared coherence — MSC) of pressure variations at ocean bottom $p$ with acceleration $a_{I_{norm}}(\eta, \kappa, t)$, corresponding to the vertical direction, for two S-net stations. Red dashed lines mark the frequencies $f_{ac}$, $f_g$, $0.1Hz$.}
    \label{fig:msc}
\end{figure}

The magnitude-squared coherence $MSC(a_{\eta, \kappa}(t), p(t))$ was computed using Welch's method with $50\%$ overlap and a segment length of $8192$ samples~\cite{18}. Each analysis was performed on a two-hour time window. The S-net network records pressure at $10 Hz$ and acceleration at $100 Hz$. In this study, the sampling frequency of the acceleration data was downsampled to $10 Hz$ by averaging to match the pressure data, as identical sampling rates are required for MSC computation. Examples of MSC spectra, corresponding to the maximum value of the coherence integral among all computed $a_{\eta, \kappa}(t)$, for stations S2N17 and S6N08 are presented in Fig.~\ref{fig:msc}.

\section{Results and Discussion}\label{sec5}

\begin{figure}[H]
    \centering
    \includegraphics[width=14cm]{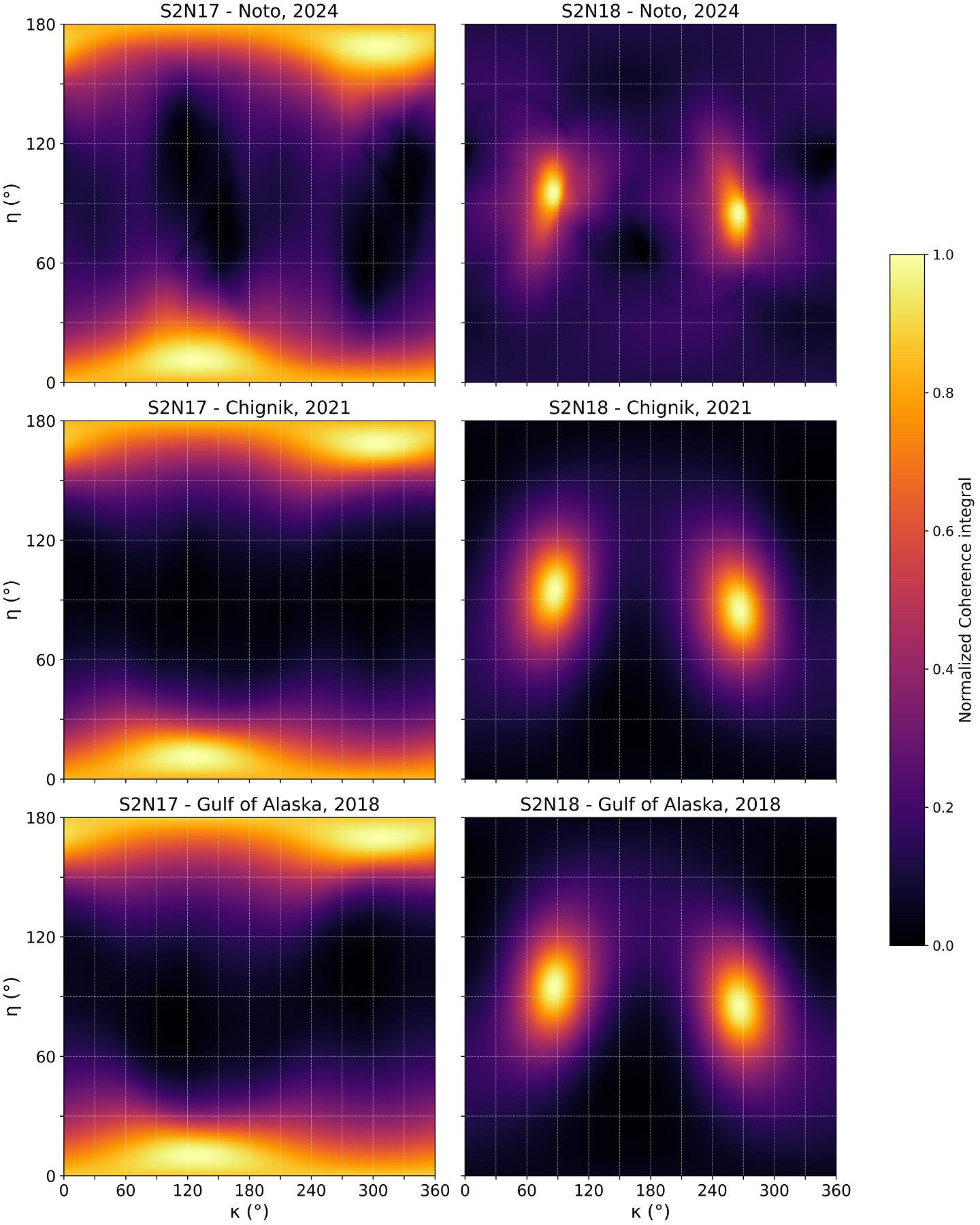}
    \caption{Coherence diagrams $I_{norm}(\eta, \kappa)$. Left column corresponds to station S2N17; right column --- S2N18. Rows from top to bottom: Noto (2024), Chignik (2021), Gulf of Alaska (2018).}
    \label{fig:d1}
\end{figure}

\begin{figure}[H]
    \centering
    \includegraphics[width=14cm]{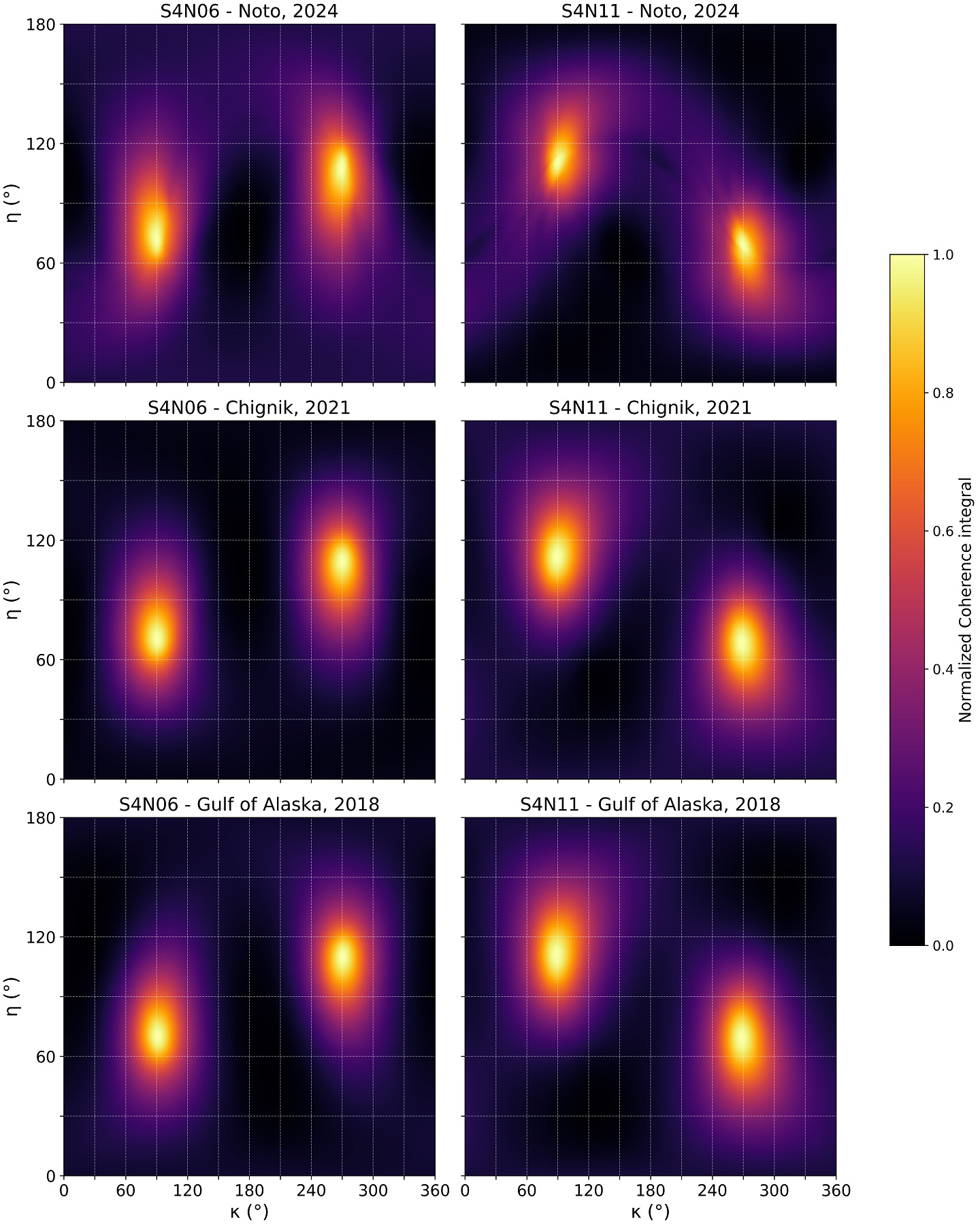}
    \caption{Coherence diagrams $I_{norm}(\eta, \kappa)$. Left column corresponds to station S4N06; right column --- S4N11. Rows from top to bottom: Noto (2024), Chignik (2021), Gulf of Alaska (2018).}
    \label{fig:d2}
\end{figure}

\begin{figure}[H]
    \centering
    \includegraphics[width=14cm]{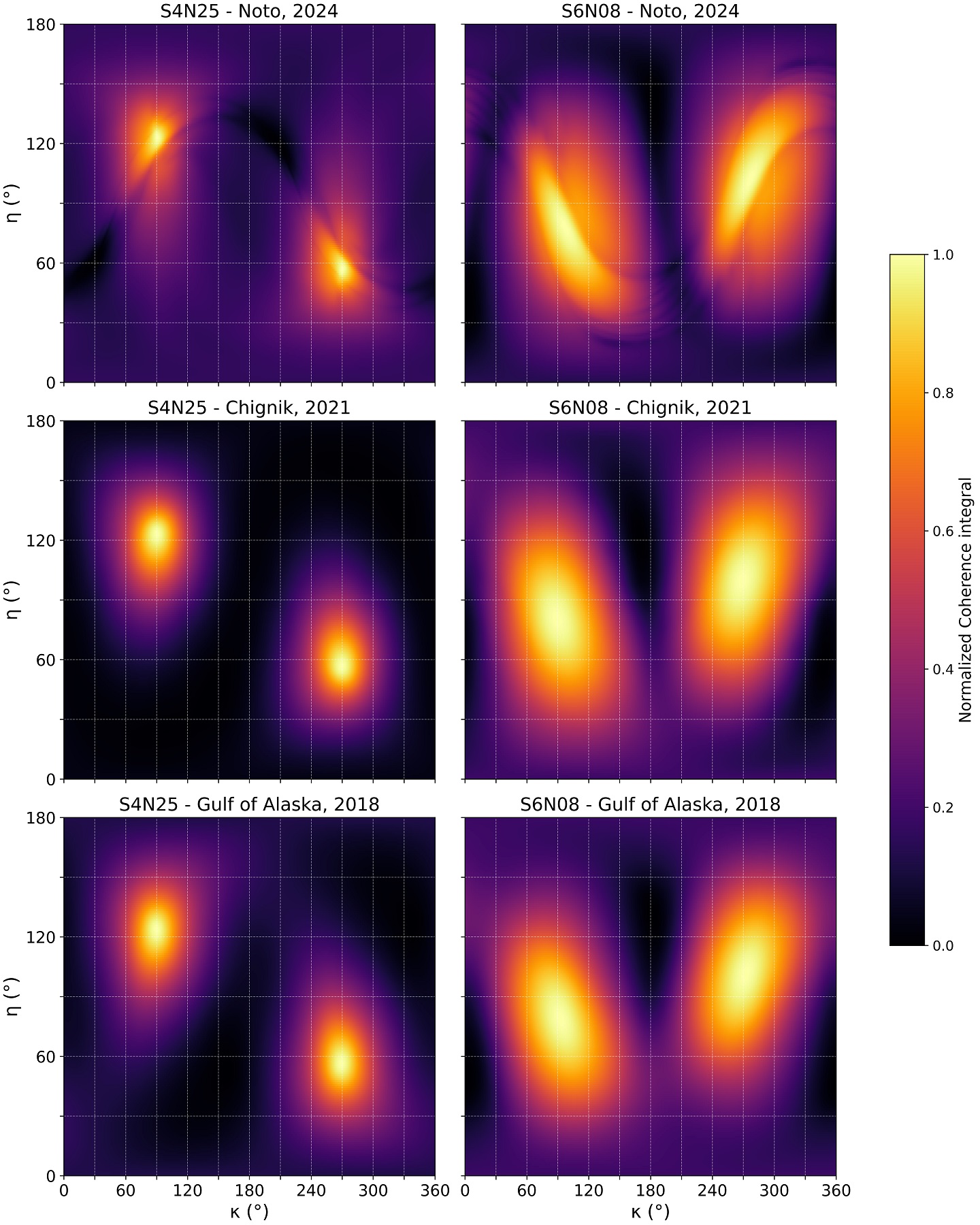}
    \caption{Coherence diagrams $I_{norm}(\eta, \kappa)$. Left column corresponds to station S4N25; right column --- S6N08. Rows from top to bottom: Noto (2024), Chignik (2021), Gulf of Alaska (2018).}
    \label{fig:d3}
\end{figure}

Using the method described in the previous section, coherence maps $I_{norm}(\eta, \kappa)$ were constructed for each station-event pair with a resolution of $1^\circ$ (Fig.~\ref{fig:d1}, \ref{fig:d2}, \ref{fig:d3}). The vertical axis of each diagram corresponds to $\eta$, and the horizontal axis corresponds to $\kappa$.  The $I_{norm}(\eta, \kappa)$ value is shown in color according to the scale depicted in the figures.

As expected, most diagrams exhibit two prominent maxima corresponding to the vertically upward and downward directions. The specific angular values $\eta, \kappa$ at these maxima are listed in Table~\ref{tab:sample}. The maxima in the diagram for station S2N17 are elongated horizontally due to a small tilt of the $z$-axis (approximately $10$ degrees) from vertical. The maxima in the diagram for station S6N08 are less distinct compared to the others, most likely due to its depth ($6 km$), whereas the other five stations lie at depths of around $2 km$. Notably, for the Noto earthquake, the maxima appear \say{smeared}, particularly for stations S4N25 and S6N08 (Fig.~\ref{fig:d3}). This may be attributed to dynamic changes in the stations’ spatial orientation during the passage of strong seismic motions. Such changes are more pronounced for the Noto event because of its proximity to the network.\par

The proposed method serves as an alternative to the gravity-based approach described in~\cite{12}, which assumes that the accelerometer records a constant component equal to the gravitational acceleration vector, oriented vertically downward. In this case, the orientation angles $\eta$ and $\kappa$ can be estimated from the following relationships, under the assumption that $\eta \in [0, 180], \kappa\in[0, 360)$:

\begin{equation}
\frac{\overline{Z}}{g} = -cos(\eta) => \eta = arccos \left( \frac{\overline{Z}}{g}\right)
\end{equation}
\begin{equation}
\frac{\overline{Y}}{\overline{X}} = tan(\kappa) => \kappa = atan2 \left(\overline{Y},\overline{X} \right) + \pi
\end{equation}
\begin{equation}
g = \sqrt{\overline{X}^2+\overline{Y}^2+\overline{Z}^2},
\end{equation}

Here, $\overline{X}$, $\overline{Y}$, and $\overline{Z}$ denote the time-averaged accelerometer readings along the $x$-, $y$-, and $z$-axes, respectively; $g$ is gravitational acceleration, and $atan2$ is the 2-argument arctangent. Table~\ref{tab:sample} presents the values of $\eta$ and $\kappa$ for the six selected stations and three earthquakes, calculated using both the cross-spectral method proposed in this work and the gravity-based method. For the cross-spectral approach, two angle pairs are listed for each station-event combination, as the method cannot distinguish between upward and downward directions (i.e., signs are ambiguous). The assignment of which maximum is labeled \say{first} or \say{second} was made arbitrarily. The angle values corresponding to the vertically upward direction are highlighted in bold.

\begin{table}[htbp]
\centering
\caption{Angular values $\eta, \kappa$ calculated using two methods}
\label{tab:sample}
\begin{tabular}{llcccccc}
\toprule
\textbf{Station} & \textbf{Year} & \multicolumn{4}{c}{\textbf{MSC-based method}} & \multicolumn{2}{c}{\textbf{g-based method}} \\
\cmidrule(lr){3-6} \cmidrule(lr){7-8}
& & $\kappa_1$ & $\eta_1$ & $\kappa_2$ & $\eta_2$ & $\kappa$ & $\eta$ \\
\midrule
\multirow{3}{*}{S2N17} & 2018 & 128 & 10 & \textbf{308} & \textbf{170} & 289 & 172 \\
                       & 2021 & 125 & 11 & \textbf{305} & \textbf{169} & 289 & 172 \\
                       & 2024 & 126 & 11 & \textbf{306} & \textbf{169} & 289 & 172 \\
\midrule
\multirow{3}{*}{S2N18} & 2018 & \textbf{86} & \textbf{95} & 266 & 85 & 86 & 95 \\
                       & 2021 & \textbf{87} & \textbf{95} & 267 & 85 & 86 & 95 \\
                       & 2024 & \textbf{86} & \textbf{95} & 266 & 85 & 86 & 95 \\
\midrule
\multirow{3}{*}{S4N06} & 2018 & \textbf{91} & \textbf{70} & 271 & 110 & 90 & 71 \\
                       & 2021 & \textbf{90} & \textbf{70} & 270 & 110 & 90 & 71 \\
                       & 2024 & \textbf{90} & \textbf{72} & 270 & 108 & 90 & 71 \\
\midrule
\multirow{3}{*}{S4N11} & 2018 & \textbf{88} & \textbf{111} & 268 & 69 & 89 & 111 \\
                       & 2021 & \textbf{89} & \textbf{112} & 269 & 68 & 89 & 111 \\
                       & 2024 & \textbf{90} & \textbf{111} & 270 & 69 & 89 & 111 \\
\midrule
\multirow{3}{*}{S4N25} & 2018 & \textbf{89} & \textbf{124} & 269 & 56 & 89 & 124 \\
                       & 2021 & \textbf{90} & \textbf{123} & 270 & 57 & 89 & 124 \\
                       & 2024 & \textbf{91} & \textbf{123} & 271 & 57 & 89 & 124 \\
\midrule
\multirow{3}{*}{S6N08} & 2018 & \textbf{92} & \textbf{79} & 272 & 101 & 91 & 80 \\
                       & 2021 & \textbf{89} & \textbf{80} & 269 & 100 & 91 & 80 \\
                       & 2024 & \textbf{97} & \textbf{78} & 277 & 102 & 91 & 80 \\
\bottomrule
\end{tabular}
\end{table}

As shown in Table~\ref{tab:sample}, no significant change in the orientation of the accelerometer axes was observed over the period under investigation. The differences between the angles obtained using the two methods generally do not exceed $1^\circ$. Significant discrepancies are observed only for stations S2N17 and S6N08. For station S2N17, the angle $\eta$ is nearly identical for both methods. The larger discrepancies in $\kappa$ can be explained by the fact that the vertical direction at this station is quite close to the $z$-axis (deviating by only $10$–$11$ degrees), resulting in small projections of the vertical onto the $x$- and $y$-axes, which amplifies numerical error. Therefore, even though the difference in $\kappa$ may appear large, the actual spatial directions defined by the angle pairs are close (for instance, if $\eta = 180$, any value of $\kappa$ represents the same direction). Station S6N08 shows the largest variation in calculated angles. It is noteworthy that the peaks in the corresponding diagrams for this station are much more diffuse than for the others. This is likely related to the station’s depth --- approximately $6 km$ --- while the others are located at about $2 km$. It is known that the ability of an earthquake to induce oscillations at a given frequency depends on its magnitude~\cite{mag}. As the depth increases, the range of forced oscillations shifts toward lower frequencies. As indicated in~\cite{11}, to excite strong oscillations at around $0.02 Hz$ (which corresponds to a depth of approximately $3600 m$), a magnitude of about $M_w8$ is sufficient. This hypothesis is supported both in the original work and in the present study. For a depth of $6000 m$ --— relevant to station S6N08 --- the magnitudes of the earthquakes analyzed are insufficient, since the lower bound of the forced oscillation frequency range is $0.015 Hz$. This is clearly illustrated in the MSC versus frequency plot (Fig.~\ref{fig:msc}). The plot for S6N08 shows that the MSC begins to decline well before the lower bound $f_g$ of the forced oscillation frequency range. For comparison, the plot for station S2N17 shows that the highest MSC values correspond to the interval $\{ f_g, min(f_{ac}, 0.1 Hz)\}$.

\section{Conclusion}\label{sec6}

This study presents a method for determining the vertical direction relative to the axes of a seismometer in a seafloor observatory using cross-spectral analysis between acceleration and pressure variations at the seafloor. A key advantage of the proposed method is that it can be applied not only to accelerometers but also to velocimeters. A natural limitation of the method is that it determines the direction only up to a sign.\par

The method was tested on six selected stations from the S-net network and validated through comparison with an independent, alternative method. For four out of the six stations, the difference between the results of the two methods was within $1^\circ$, consistent with the angular resolution of the grid.\par

The process of determining the accelerometer’s orientation involves constructing a diagram, which itself serves as a valuable source of information. For the Noto earthquake, the peaks in the diagrams appeared smeared, most likely due to changes in the orientation of the accelerometer axes during the event, induced by the large amplitude of seafloor motion caused by the nearby earthquake.\par

No significant change in the orientation of the accelerometer axes was observed for the selected stations during the period from 2018 to 2024.\par

The least distinct peaks in the diagrams --- and consequently, the greatest uncertainty in the determination of the accelerometer orientation --- were observed for station S6N08, the deepest among those considered ($6 km$). This is attributed to the shift of the forced oscillation range toward lower frequencies with increasing depth. It was shown that earthquakes with magnitudes in the range $M_w7.5$–$8$ are insufficient to effectively excite oscillations across the full range of forced frequencies at a depth of $6000 m$: $[0.015, 0.065] Hz$. For stations at depths of approximately $2000 m$, the relevant frequency band was sufficiently excited to enable determination of the accelerometer orientation with an accuracy within $1^\circ$. The only exception was station S2N17, but in this case, the large error in the calculated angles $\eta, \kappa$ corresponds to a small actual deviation in solid angle, since the vertical direction was close to the accelerometer’s $z$-axis.

\backmatter

\bmhead{Supplementary information}

The supplemental materials include bathymetry and gradient magnitudes near each of the considered stations, as well as cross-spectra (magnitude-squared coherence — MSC) of pressure variations at the ocean bottom with computed vertical accelerations of seismic oscillations of the ocean bottom registered by S-net stations.

\bmhead{Acknowledgements}

The authors sincerely appreciate National Research Institute for Earth Science and Disaster Resilience (NIED) for providing S-net data. We thank GEBCO Compilation Group for bathymetry data.

\bmhead{Funding}
The study was conducted under the state assignment of Lomonosov Moscow State University. 

\bmhead{Data availability}
The NIED S-net data used in this study is available at https://doi.org/10.17598/nied.0007. The bathymetry is available at https://doi.org/10.5285/f98b053b-0cbc-6c23-e053-6c86abc0af7b.

\bmhead{Author Contributions}

Conceptualization: NMA; methodology: POV, KSV; software: POV; formal analysis and investigation: POV, KSV; visualization: POV; discussion on the results: POV, KSV, NMA; writing the original draft: POV; review the draft: NMA, KSV; editing the final version of the paper: POV. All authors approved the final version of the paper.

\bibliography{sn-bibliography}

\end{document}